\documentstyle[prd,aps,epsf,epsfig]{revtex} 
\begin{document}
\draft
\twocolumn[\hsize\textwidth\columnwidth\hsize\csname@twocolumnfalse\endcsname
%
\title{ Spatial stochastic resonance in 1D Ising systems}

\author
{Z. N\'eda, \'A. Rusz and E. Ravasz}
\address{Babe\c{s}-Bolyai University, 
         Dept. of Theoretical Physics \\
         str. Kogalniceanu 1, RO-3400, Cluj, ROMANIA}
\author{Porus Lakdawala}
\address{Homi Bhabha Center for Science Education \\
V. N. Purav Marg, Mumbai 400088, India}
\author{P.M. Gade}
\address{Academia Sinica, Institute of Physics \\
         Nankang 11592, Taipei, R.O.C.}

\maketitle
\centerline{\small (Last revised \today)}

\begin{abstract}

The 1D Ising model is analytically studied in a spatially periodic and 
oscillatory external
magnetic field using the transfer-matrix method.  For low enough 
magnetic field intensities the correlation between 
the external magnetic field and the response in magnetization presents
a maximum for a given temperature. The phenomenon can be interpreted as
a resonance phenomenon induced by the stochastic heatbath. This novel
"spatial stochastic resonance" has a different origin than the classical
stochastic resonance phenomenon.

\end{abstract}

\vspace{2pc}
]

\vspace{1cm}


\section{Introduction}

Many recent papers 
\cite{neda1,neda2,leung1,leung2,brey,schimansky1,schimansky2,sides1,sides2}
revealed unequivocally 
the phenomenon of stochastic resonance (SR) \cite{sr} in the kinetic
Ising model driven by a temporary oscillating magnetic field. 
SR was anticipated by
considering the Ising model as a system of coupled two-state oscillators in a
stochastic force-field which is taken to be thermal fluctuations. 
In this sense the system has all the ingredients necessary to observe
the classical phenomenon of stochastic resonance.

In the present paper we intend to study the 1D ferromagnetic Ising model in a
spatially periodic and oscillatory $B(i)$  external magnetic field
($B(i+\lambda)=B(i)$, $<B(i)>_i=0$, $2\lambda$ the spatial period of  
the field, and $i$ the lattice points $i=1,2,3,......2p\lambda$, 
$p=1,2,.....$).
We impose periodic boundary conditions, thus 
$S(2p\lambda+1)=S(1)$.
The Hamiltonian of the system writes as
\begin{equation}
H=-J\sum_{i=1}^{2p\lambda} S(i) S(i+1)- \mu \sum_{i=1}^{2p\lambda} B(i) S(i),
\end{equation}
($\mu$ the magnetic moment of the $S(i)=\pm 1$ Ising spins).
The magnetic field is taken 
stationary in time. 

For the proposed problem is easy to realize that at $T=0$ thermodynamic
temperature and not to high magnetic field intensities,
due to the oscillatory nature of the external magnetic field the
$\sigma=<B(i) S(i)>$ (the brackets denotes both a spatial and ensemble 
average)
correlation is greatly reduced, 
i.e. the infinite correlation
length ($\xi(0)=\infty$) between the spins 
competes with the finite $2\lambda$ period of $B(i)$.
For $T=\infty$ is also obvious that due to the leading stochastic
contribution $\sigma=0$. We expect that for a given
finite temperature the $\xi$ correlation length will be of the order of
the $\lambda$ period of $B(i)$ and
thus the $\sigma$ correlation will reach a 
maximal value. This resonance like phenomenon is  induced
by the stochastic force field (temperature) for the energetically frustrated
system at $T=0$. Therefore, in order to distinguish this phenomenon from  the 
classical SR phenomenon we named it 
{\em spatial stochastic resonance }.

\section{The method}

To give an exact solution for the proposed 
problem we choose the most simplest possible $B(i)$ configuration
with the above imposed properties. We choose
$B(i)=B$ for $i=2n\lambda+1$ ($n=0,1,2....p-1$), $B(i)=-B$ for
$i=(2n+1)\lambda+1$, and B(i)=0 for all other lattice points.
We are interested in the $<S(1)>$ average magnetization at
the $i=1$ position from where
the $\sigma$ correlation is easily determined.
From the chosen magnetization profile we get:
\begin{equation}
\sigma=<B(i) S(i)>=pB (<S(1)>-<S(\lambda+1)>)
\end{equation} 
From symmetry arguments $<S(1)>=-<S(\lambda+1)>$, and  we can write: 
\begin{equation}
\sigma_p=2pB<S(1)>
\label{sigma}
\end{equation}

In order to determine $<S(1)>$ we calculate i.) the 
$Z_{2p\lambda}$ partition function of the system ii.) the
$Z_{2p\lambda}^+$ partition function for $S(1)=1$ imposed 
condition, iii.) and the $Z_{2p\lambda}^-$ partition function
for the $S(1)=-1$ imposed condition.
We get the desired $<S(1)>$ value, as:
\begin{equation}
<S(1)>=\frac{Z_{2p\lambda}^+-Z_{2p\lambda}^-}{Z_{2p\lambda}}
\label{av}
\end{equation}

During our calculations we use several matrices whose explicit
forms are given in the Appendix.

\section{The $L=2\lambda$ length chain $(p=1)$  }

With the notations $j=J/kT$ and $h=\mu B/kT$ ($T$ the temperature of the
system, $k$ the Boltzmann constant) the partition function 
$Z_{2\lambda}$ writes as
\begin{equation}
Z_{2\lambda}=\sum_{S(1)} \sum_{S(2)} ... \sum_{S(2\lambda)}
e^{j[\sum_{i=1}^{2\lambda}S(i) S(i+1)]+hS(1)-hS(\lambda+1)},
\end{equation}
(the sums are for $S(i)=\pm 1$).

We use the transfer-matrix method to calculate $Z_{2\lambda}$
\begin{eqnarray}
& \nonumber Z_{2\lambda}=\sum_{S(1)} \sum_{S(2)} ... \sum_{S(2\lambda)}
<S(1)\mid I_+ \mid S(2)> \\ 
& \nonumber <S(2) \mid I_o \mid S(3)>... 
<S(\lambda) \mid I_o \mid S(\lambda+1) >   \\ 
&  \nonumber <S(\lambda+1) \mid I_-
\mid S(\lambda+2) > <S(\lambda+2) \mid I_o \mid S(\lambda+3) >\\
& \nonumber... <S(2\lambda-1 \mid I_o \mid S(2\lambda) >
< S(2\lambda) \mid I_o \mid
S(1) > 
\end{eqnarray}
with the $I_o$ and $I_{\pm}$  matrices given in the Appendix (\ref{Io}). 
It is now immediate to realize that,
\begin{equation}
Z_{2\lambda}=Tr(I_+ I_o^{\lambda-1} I_- I_o^{\lambda-1})
\end{equation}
(where $Tr(A)$ denotes the trace of the $A$ matrix).
Introducing the $M$ (\ref{M}) diagonal matrix we get 
\begin{eqnarray}
I_+=MI_o \\
I_-=M^{-1} I_o
\label{rel}\\
Z_{2\lambda}=Tr(W)\\
W=MI_o^{\lambda}M^{-1}I_o^{\lambda}
\end{eqnarray}
Choosing a representation where $I_o$ becomes diagonal we get:
\begin{eqnarray}
I_o'=UI_oU^{-1} \\
M'=UMU^{-1} \\
W'=M' I_o'^{\lambda} M'^{-1} I_o'^{\lambda}\\
Z_{2\lambda}=Tr(W')
\end{eqnarray}
(the $U$, $I_o'$ and $M'$ matrices are also given in the Appendix).
After some  elementary algebra one will find:
\begin{eqnarray}
& \nonumber Z_{2\lambda}=2^{2\lambda} \{cosh^2(h)[cosh^{2\lambda}(j)+sinh^{2\lambda}(j)] \\
& -2sinh^2(h)sinh^{\lambda}(j) cosh^{\lambda} (j)]\}
\end{eqnarray}
For $Z_{2\lambda}^+$ and $Z_{2\lambda}^-$
\begin{equation}
Z_{2\lambda}^{\pm}=\sum_{S(2)} \sum_{S(3)} ... \sum_{S(2\lambda)}
e^{j [\sum_{i=1}^{2\lambda} S(i) S(i+1)] + h S(1)-h S(\lambda+1)} 
\end{equation}
(for the $\pm$ cases we have $S(1)=\pm 1$ respectively) 
we perform a similar calculation:
\begin{eqnarray}
& \nonumber Z_{2\lambda}^{\pm}=\sum_{S(2)} \sum_{S(3)} ... \sum_{S(2\lambda)} 
<S(2\lambda) \mid S_{\pm} \mid S(2)> \\ 
& \nonumber <S(2) \mid I_o \mid S(3) > ... 
<S(\lambda) \mid I_o \mid S(\lambda+1)> \\ 
& \nonumber <S(\lambda+1) 
\mid I_- \mid
 S(\lambda+2)><S(\lambda+3) \mid I_o \mid S(\lambda+4)>\\
& \nonumber ... <S(2\lambda-1) \mid I_o \mid S(2\lambda)>
\end{eqnarray}
\begin{equation} 
Z_{2\lambda}^{\pm}= Tr(S_{\pm} I_o^{\lambda-1}
I_- I_o^{\lambda-2} )
\end{equation}
The $S_{\pm}$ matrices are also given in the Appendix (\ref{S}).
Using (\ref{rel}) we can write: 
\begin{eqnarray}
Z_{2\lambda}^{\pm}=Tr(P_{\pm} I_o^{\lambda} M^{-1} I_o^{\lambda}) \\
P_{\pm}=I_o^{-1} S_{\pm} I_o^{-1} 
\end{eqnarray}
Again, we calculate the trace in the representation where $I_o$ is diagonal
and we get:
\begin{eqnarray}
& \nonumber Z_{2\lambda}^{\pm}=2^{2\lambda-1} 
e^{\pm h} \{cosh(h)[cosh^{2\lambda}(j)+
sinh^{2 \lambda} (j)] \mp  \\
& 2 sinh(h) sinh^{\lambda}(j) cosh^{\lambda}(j)\}
\end{eqnarray}
With the obtained $Z_{2\lambda}$ and $Z_{2\lambda}^{\pm}$ values 
$<S(1)>$ is easily determined (\ref{av}) and for the $\sigma_p$ 
(\ref{sigma}) correlation we
get finally:
\begin{equation}
\sigma_1=2B\frac{tanh(2h)}{1+\frac{1}{cosh(2h)} (\frac{1+tanh^{\lambda}(j)}
{1-tanh^{\lambda}(j)})^2} 
\label{fin1}
\end{equation}

\section{The $L=2p\lambda$ ($p>1$) length chain}

In order to be able to make the calculations easily in the $p>1$
case we first write $Z_{2\lambda}^{\pm}$ in a more 
convenient form. 
\begin{eqnarray}
& Z_{2\lambda}^{\pm}=Tr(P_{\pm}' I_o'^{\lambda} M'^{-1} I_o'^{\lambda})=
Tr(R_{\pm}' W') \\ 
& R_{\pm}'=P_{\pm}' M'^{-1}
\end{eqnarray}  
Is easy to realize
that:
\begin{eqnarray}
Z_{2\lambda}=Z_{2\lambda}^+ + Z_{2\lambda}^-=Tr(W') \\
Z_{2\lambda}^+-Z_{2\lambda}^-=Tr(R' W') 
\end{eqnarray}
For the $p>1$ case is immediate that:
\begin{equation}
Z_{2p\lambda}=Tr(W'^p)
\end{equation}
Writing up the effective forms of $Z_{2p\lambda}^{\pm}$ like in the
$p=1$ case, one can also show  that:
\begin{equation}
Z_{2p\lambda}^{\pm}=Tr(R_{\pm}' W'^p)
\end{equation}
In the representation where $W'$ is diagonal it is easy now to
calculate $<S(1)>$ and $\sigma_p$. Denoting  
by $\chi_1$ and $\chi_2$ ($\chi_1\ge \chi_2$) the eigenvalues 
of $W'$, after some simple algebra we find:
\begin{eqnarray}
& \sigma_p=2pB\frac{Tr(R' W'^p)}{Tr(W'^p)}=\sigma_1 \frac{Tr(W')}
 {\sqrt{\Delta}}\frac{\chi_1^p-\chi_2^p}{\chi_1^p+\chi_2^p} \\
& \Delta=Tr(W')^2-4 Det(W') \\
& \chi_{1,2}=\frac{Tr(W')\pm \sqrt{\Delta} } {2} 
\end{eqnarray}
In the limit $p\rightarrow \infty$ we get the simple formula
\begin{equation}
\sigma_{\infty}=\sigma_1 \frac{Tr(W')} {\sqrt(\Delta)},
\label{fin2}
\end{equation}   
which is easily computable from the (\ref{W}) $W'$ matrix given
in the Appendix.

\section{Discussion}

Equation (\ref{fin1}) and (\ref{fin2}) gives us the 
$\sigma=<B(i) S(i)>$ correlations
for the $p=1$ and $p=\infty$ periodic chains. The $\sigma(T)$ correlation
is proportional with the $<S(1)>(T)$ curves (\ref{sigma}). On Fig.~1 we plotted
$<S(1)>(T)$ for three different applied magnetic field
intensities.

As expected, for $\mu B/J <2$ (when the interaction with the external 
field is weaker
than the interaction with the neighboring spins) a clear resonance
like behavior is obtained. Both for $p=1$ and $p=\infty$ $<S(1)>(T)$
exhibits a clear maximum at a $T_r\ne0$ resonance temperature.   
It is also observable that the $<S(1)>(T)$ curves for $p=1$ and
$p=\infty$ are very close, thus the $p=1$ result is qualitatively 
well describing the $p>1$ cases as well.
The $T_r$ resonance temperature depends both on the $B$ intensity
of the applied magnetic field and the characteristic $\lambda$ distance 
of the spatial oscillations of $B(i)$. On Fig.~2 we illustrate the
$T_r(B)$ dependence, and on Fig.~3 the $T_r(\lambda)$ trend.
From Fig.~2 we learn that in the $B\rightarrow 0$ limit the $T_r$
values are converging to a constant (which is dependent on $p$), and
in the $\mu B/J\ge 2$ limit $T_r=0$, thus  no resonance 
behavior is obtained.  The $T_r(\lambda)$ variations (Fig.~3) are also
the one expected from our phenomenological considerations. 
In the limit $\lambda\rightarrow \infty$ we get $T_r\rightarrow 0$, and
$T_r$ is monotonically decreasing with increasing $\lambda$ values.
It is interesting to note that for the minimal possible $\lambda$ value
($\lambda=1$) the resonance like behavior is still present but 
the $p=1$ and $p=\infty$ curves are much more distant comparable with
large $\lambda$ values case.

\section{Conclusions}

In the present paper we studied the response of 1D Ising chain
to a spatially periodic and oscillatory magnetic field. Considering
the most simple magnetic field profile we exactly solved the problem
by using the transfer-matrix method. We found that the 
$\sigma=<B(i) S(i)>$ correlation between the applied magnetic field and
the local magnetization exhibits a maximum for a given $T_r$ resonance
temperature (Fig.~1). The $T_r$ resonance 
temperature depends monotonically on the
spatial oscillation length of the
magnetic field (Fig.~3). The value of $T_r$ depends also on the $B$ intensity
of the magnetic field, and becomes independent of $B$ in the small
$B$ values limit (Fig.~2). The length of the chain $L=2p\lambda$, characterized
by the $p$ integer value has small influence on the observed
resonance-like behavior for large $\lambda$ values.  
The obtained resonance-like behavior is induced by the stochastic
heat-bath and has a different origin from the classical stochastic
resonance phenomenon. We named it {\em spatial stochastic resonance}.  

\section{Appendix}

\begin{eqnarray}
& I_o=\left[ \begin{array}{cr}
e^j      &  e^{-j} \\
e^{-j}   &  e^j  
\end{array} \right]; \:
I_{\pm}=\left[ \begin{array}{cr}
e^{j \pm h}  &  e^{-j \pm h} \\
e^{-j \mp h} &  e^{j \mp h} 
\end{array} \right]  
\label{Io} \\
& M=\left[ \begin{array}{cr}
e^h & 0 \\
0   & e^{-h} 
\end{array} \right]; \:
U=\frac{1}{\sqrt{2}} \left[ \begin{array}{cr}
1  &  1 \\
1  &  -1
\end{array} \right]; 
\label{M} \\
& M'=\left[ \begin{array}{cr}
cosh(h) & sinh(h) \\
sinh(h) & cosh(h)
\end{array} \right]; \\
& M'^{-1}=\left[ \begin{array}{cr}
cosh(h)  &  -sinh(h) \\
-sinh(h) &  cosh(h)
\end{array} \right] \\ 
& I_o'=2 \left[ \begin{array}{cr}
cosh(j) &  0  \\
0       &  sinh(j) 
\end{array} \right] ; \:
S_{\pm}= e^{\pm h}\left[ \begin{array}{cr}
e^{\pm 2j}  &  1 \\
1       &  e^{\mp 2j} 
\end{array} \right] 
\label{S} \\
& W'=2^{2\lambda} \left[ \begin{array}{cr}
w'_{11}   &   w'_{12} \\
w'_{21}   &   w'_{22}
\end{array} \right] 
\label{W} \\
& \nonumber w'_{11}=
cosh^2(h)cosh^{2\lambda}(j)-sinh^2(h)cosh^{\lambda}(j)sinh^{\lambda}(j) \\
& \nonumber w'_{12}=sinh(h) cosh(h)sinh^{\lambda}(j)[sinh^{\lambda}(j)-cosh^{\lambda}(j)  \\
& \nonumber w'_{21}=sinh(h)cosh(h)cosh^{\lambda}(j)[cosh^{\lambda}(j)-sinh^{\lambda}(j)]    \\ 
& \nonumber w'_{22}=cosh^2(h)sinh^{2\lambda}(j)-sinh^2(h)cosh^{\lambda}(j)sinh^{\lambda}(j) 
\\
& P_{\pm}'=\frac{e^{\pm h}}{2} \left[ \begin{array}{cr}
1       &  \pm 1  \\
\pm 1   &   1
\end{array} \right]; \\
& R'=\left[ \begin{array}{cr}
0  &  1 \\
1  &  0
\end{array} \right]; \:
R'_{\pm}=\frac{1}{2} \left[ \begin{array}{cr}
1      &  \pm 1 \\
\pm 1  &  1 
\end{array} \right]
\end{eqnarray}


\vspace{1cm}

{\bf Figures Captions:}

\vspace{0.5cm}

{\bf Fig.~1} Characteristic shape of $<S(1)>(T)$ 
for three different magnetic field intensities ($ \mu B/k=0.1; 1.0; 2.1$).
The continuous lines are for $p=1$, the dashed ones for $p=\infty$,
$\lambda=20$ for all curves.

{\bf Fig.~2} The $T_r$ resonance temperature as a function of the
applied magnetic field intensity. We draw the results for two different 
$\lambda$ values. The continuous lines are for $p=1$, the dashed
ones for $p=\infty$.

{\bf Fig.~3} The $T_r$ resonance temperature as a function of the
$\lambda$ length. We draw the results for two different applied
magnetic field intensities. The continuous lines are for $p=1$, the
dashed ones for $p=\infty$.

\onecolumn


\begin{figure}[htp]
\epsfig{figure=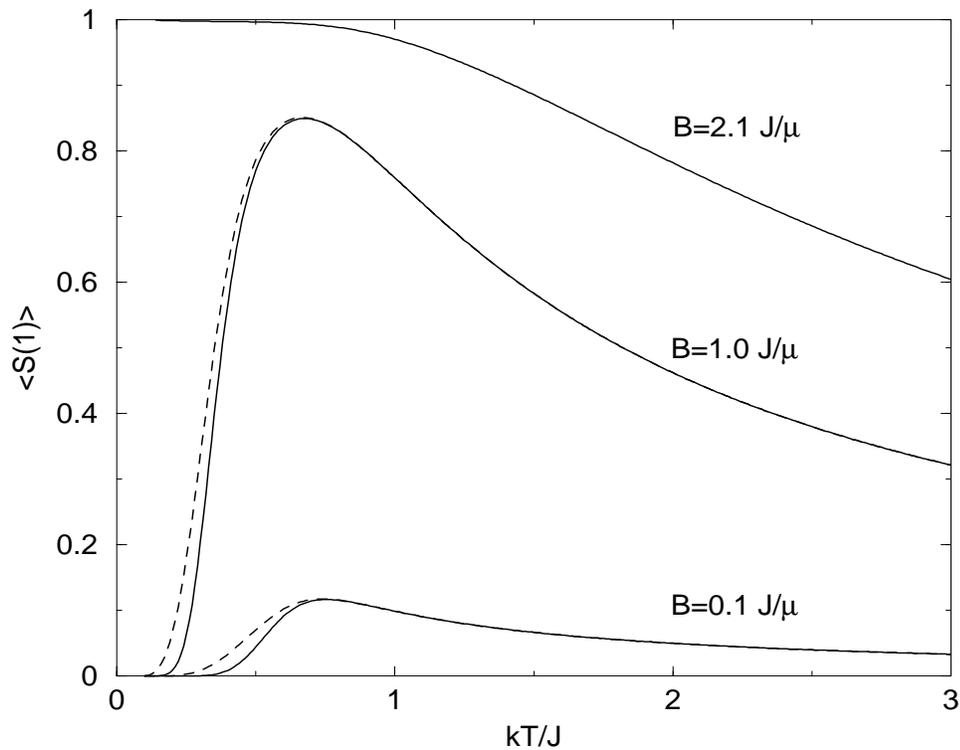,height=4.0in,width=5.0in,angle=-0}
\caption{Characteristic shape of $<S(1)>(T)$ 
for three different magnetic field intensities ($\mu B/k=0.1; 1.0;2.1$).
The continuous lines are for $p=1$, the dashed ones for $p=\infty$, 
$\lambda=20$ for all curves. }
\label{fig1}
\end{figure}


\begin{figure}[htp]
\epsfig{figure=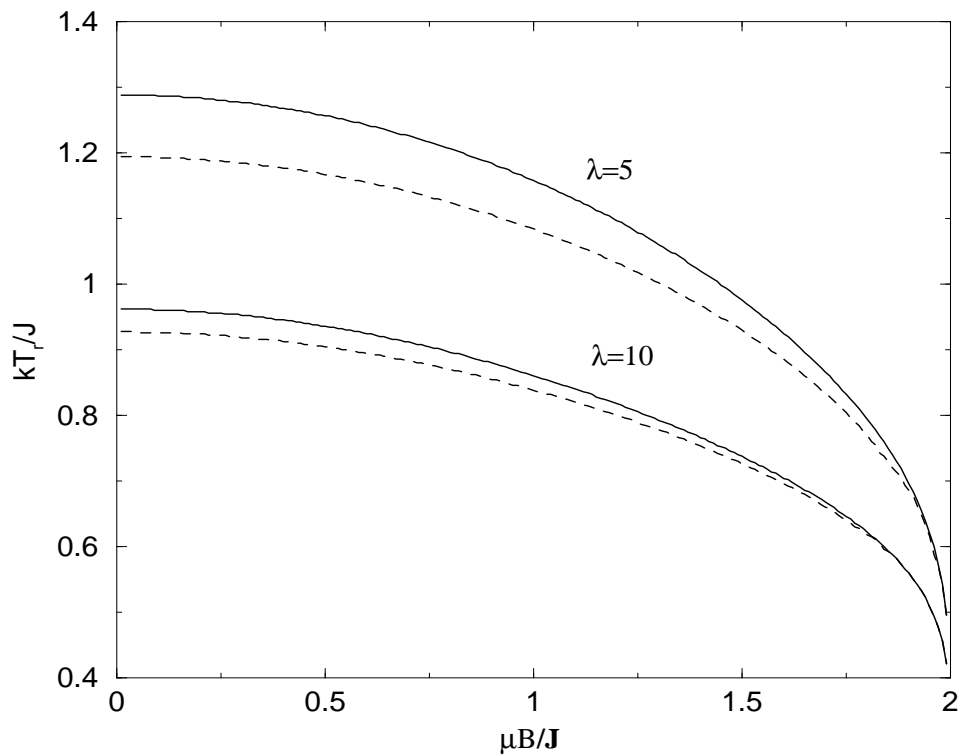,height=4in,width=5.0in,angle=-0}
\caption{The $T_R$ resonance temperature as a function of the
applied magnetic field intensity. We draw the results for two different
$\lambda$ values. The continuous lines are for $p=1$, the dashed ones
for $p=\infty$. 
}
\label{fig2}
\end{figure}


\begin{figure}[htp]
\epsfig{figure=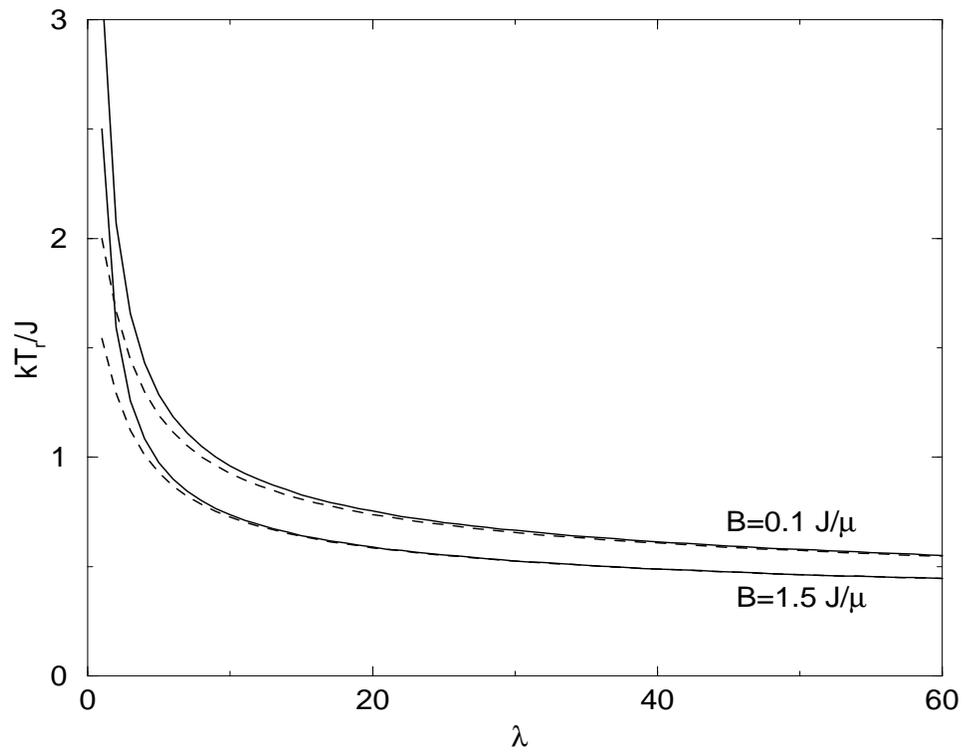,height=4.0in,width=5.0in,angle=-0}
\caption{ The $T_r$ resonance temperature as a function of the
$\lambda$ length. We present results for two different
applied magnetic field intensities. The continuous lines are for
$p=1$, the dashed ones for $p=\infty$.
}
\label{fig3}
\end{figure}

\end{document}